\documentclass{PoS}
\usepackage{colordvi}
\usepackage{graphicx}
\usepackage{epsfig}
\usepackage{rotating}

\def\bra#1#2{\ifx#2\ket\langle#1\else\langle#1\vert\fi#2}
\def\ket#1{\vert#1\rangle}

\leftline{\parbox{2.8cm}{\large\rm DESY~08-134 ~~HU-EP-08/40 ~~MS-TP-08-20 ~~SFB/CPP-08-77}
\vspace{1cm}}
\title{The finite-temperature phase structure \\ of lattice QCD with twisted-mass Wilson fermions }

\ShortTitle{Finite-temperature phase structure of QCD with twisted-mass Wilson fermions }

\author{The tmfT Collaboration:}
\author{
  \speaker{Ernst-Michael~Ilgenfritz}%
  \thanks{Supported by DFG via Forschergruppe Gitter-Hadronen-Ph\"anomenologie FOR 465.}\\
  Institut f\"ur Physik, Humboldt-Universit\"at zu Berlin, Newtonstr. 15, D-12489 Berlin, Germany \\
  Institut f\"ur Physik, Karl-Franzens-Universit\"at Graz, Universit\"atsplatz 5, A-8010 Graz, Austria \\
  E-mail: \email{ilgenfri@physik.hu-berlin.de}
}
\author{
  Karl Jansen \\
  DESY Zeuthen, Platanenallee 6, D-15738 Zeuthen, Germany \\ 
  E-mail: \email{Karl.Jansen@desy.de}
}
\author{
  Maria Paola~Lombardo \\
  Laboratori Nazionali di Frascati, INFN, via E. Fermi 40, I-100044 Frascati, Roma, Italy \\
  E-mail: \email{Mariapaola.Lombardo@lnf.infn.it}
}
\author{
  Michael~M\"uller-Preussker and Marcus~Petschlies \\
  Institut f\"ur Physik, Humboldt-Universit\"at zu Berlin, Newtonstr. 15, D-12489 Berlin, Germany \\
  E-mail: \email{mmp@physik.hu-berlin.de},\\ 
\hspace{1.1cm}  \email{Marcus.Petschlies@physik.hu-berlin.de}
}
\author{
  Owe~Philipsen and Lars~Zeidlewicz \\ 
  Institut f\"ur Theorische Physik, Westf\"alische Wilhems-Universit\"at M\"unster, \\ Wilhelm-Klemm-Str. 9, D-48149 M\"unster, Germany, \\ 
  E-mail: \email{o.philipsen@uni-muenster.de},
          \email{zeidlewicz@uni-muenster.de}
}

\abstract{
  We report progress in our  
  exploration of the finite-temperature phase structure 
  of two-flavour lattice QCD with twisted-mass Wilson fermions and a 
  tree-level Symanzik-improved gauge action for a temporal lattice size 
  $N_{\tau}=8$. Extending our investigations to a wider region of 
  parameter space we gain a global view of the rich phase structure.
  We identify the finite temperature transition/crossover 
  for a non-vanishing twisted-mass parameter in the neighbourhood of the 
  zero-temperature critical line at sufficiently high $\beta$.
  Our findings are consistent with Creutz's conjecture of a conical 
  shape of the finite temperature transition surface. 
  Comparing with NLO lattice $\chi{\rm PT}$ we achieve an improved 
  understanding of this shape.
}

\FullConference{The XXVI International Symposium on Lattice Field Theory\\
                 July 14 - 19, 2008\\ Williamsburg, VA, USA}

\graphicspath{{./Figures/}}

\begin{document}

\vspace{-0.1cm}
\section{Introduction}
\label{sec:introduction}
\vspace{-0.1cm}
The goal of the tmfT collaboration~\cite{Ilgenfritz:2006tz} is to explore the 
applicability of the 2-flavour 
Wilson-twisted-mass (Wtm) fermion formulation set-up as described in 
Ref.~\cite{Shindler:2007vp}  
for investigations of lattice thermodynamics.
We refer to~\cite{Shindler:2007vp} for the definition of the Wtm action and 
its parameters $\kappa$ and $\mu$. 
The staggered-fermion formulation is computationally inexpensive~\cite{Jansen:2008},
however conceptually controversial~\cite{Creutz:2007rk}.
On the other hand, the non-perturbatively improved Wilson fermion formulation 
requires the calculation of the improvement coefficients and furthermore needs 
operator improvement. 
Thus, the Wtm formulation, nowadays combined with a tree-level Symanzik-improved gauge 
action~\cite{Boucaud:2008xu}, appears as an appealing alternative for finite 
temperature lattice simulations, whose potential should be explored. It offers automatic 
$\mathcal{O}(a)-$ improvement by 
tuning the bare, untwisted quark mass only.
This makes high-statistics simulations for thermodynamical problems affordable with 
pseudoscalar masses as low as $m_{\pi} \gtrsim 300\,{\rm MeV}$~\cite{Boucaud:2007uk}.

As a necessary preparatory step we characterise the phase structure of the model by 
locating the transition/crossover lines and surfaces 
in the $\beta-\kappa-\mu-$coupling
space. In contrast to our report at Lattice 2007~\cite{Ilgenfritz:2007qr}, 
we now find clear evidence for the Aoki phase at $\beta \leq 3$,
lower than searched for previously, and a first order phase 
transition surface in an intermediate $\beta$ region beginning with 
$\beta \approx 3.4$, consistent with earlier findings.
For larger couplings $3.6 \le \beta \le 3.8$, we observe a
thermal transition at $\kappa_T(T\neq 0)$ near $\kappa_c(T=0)$, which 
emanates from the first order like structure at smaller $\beta$-values.
With growing $\beta$, it splits up to become a succession  
of two transitions 
confinement$\rightarrow$deconfinement$\rightarrow$confinement in 
the $\kappa$-direction.
The confinement aspect of these transitions 
is exhibited by the Polyakov loop and its susceptibility,
the chiral aspect by the pion norm (the integrated pseudoscalar correlator)
and the scalar and pseudoscalar condensates.

Our ultimate goal is to investigate the finite temperature transition (or crossover) 
at maximal twist and physically relevant pion masses. The most obvious procedure
of directly scanning the transition at maximal twist has proven the hardest to 
tackle, since its precise location is yet unknown. 
Still, our efforts so far gave us more detailed insights than 
reported in~\cite{Ilgenfritz:2007qr} into the phase structure at non-vanishing 
twist parameter $\mu$ for a larger range of $\beta$-values.
Our findings are qualitatively
consistent with predictions from $\chi{\rm PT}$ and continuum 
symmetry arguments~\cite{Creutz:2007fe}.

\vspace{-0.1cm}
\section{Preview of the phase structure}
\label{sec:preview}
\begin{figure}[h]
\begin{center}
\begin{minipage}[ht]{0.51\textwidth}
\includegraphics[width=0.95\textwidth]{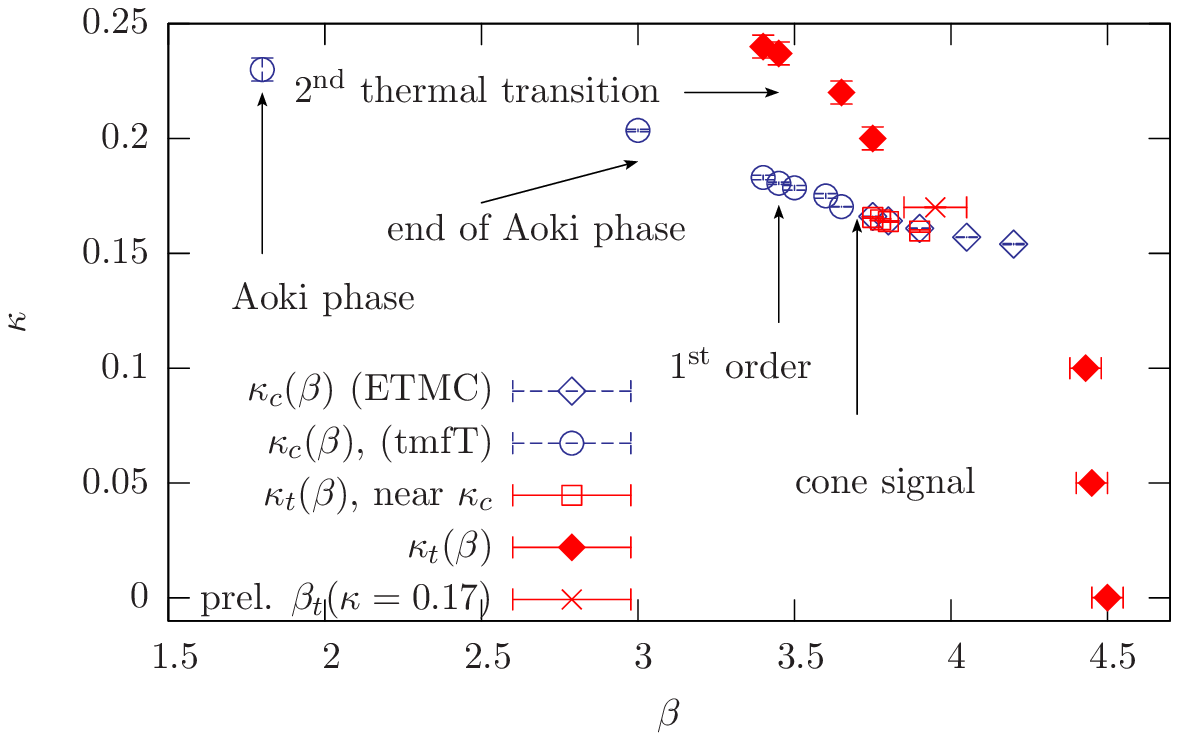}
\end{minipage}
\begin{minipage}[ht]{0.48\textwidth}
  \includegraphics[width=0.75\textwidth,angle=-90]{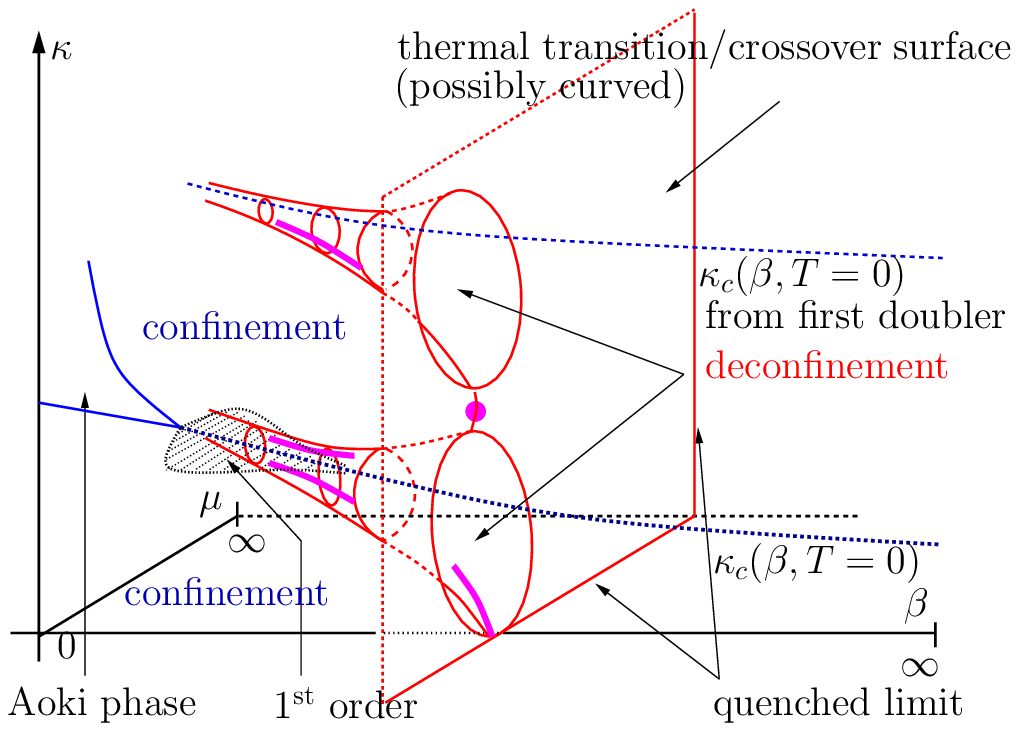}
\end{minipage}
\end{center}
\vspace{-0.6cm}
\caption{{\bf Left:} Map of our simulations performed at $\mu<0.007$. 
{\bf Right:} Emerging phase structure; regions identified by
tmfT measurements are marked by magenta symbols.}
\label{fig:overview}
\end{figure}

\vspace{-0.1cm}
There are three sources for our expectations concerning the phase structure of 
2-flavour Wtm LQCD at finite temperature: 
(i) The phase structure at zero temperature 
was predicted using $\chi{\rm PT}$~\cite{Creutz:1996bg,Sharpe:2004ps,Munster:2004am} 
and subsequently verified numerically~\cite{Farchioni:2004us}; the parameter space 
at $T=0$ was shown to contain a phase of broken parity-flavour symmetry 
(Aoki-scenario) in the $\beta-\kappa-$plane at strong 
coupling~\cite{Ilgenfritz:2003gw} (for Wilson gauge action)
and a first order phase transition surface in 
the $\beta-\mu-$plane (the first order scenario of Sharpe and Singleton)
adjacent to the former and extending into the weak coupling region. Both scenarios 
crucially depend on the choice of fermion and gauge 
action~\cite{Farchioni:2004fs}.
(ii) The general properties of the phase structure for (non-) perturbatively 
improved Wilson fermions, as established 
by e.g. the CP-PACS~\cite{Ukita:2006pc} and DIK~\cite{Bornyakov:2007zu} 
collaborations,
should be found in the $\beta-\kappa-$plane at weaker coupling, too. 
(iii) Based on the argument that in the continuum theory the physical transition 
temperature cannot depend on the twist angle, Creutz~\cite{Creutz:2007fe} conjectured 
the finite temperature phase transition/crossover surface to be closed and form a cone
(cf. the right panel of Fig.~\ref{fig:overview})
around the critical line $\kappa_c(\beta)$ for a range of $\beta$-values somewhere 
between the Aoki phase and the deconfinement transition in the quenched limit 
($\beta_{\rm dec} \approx 4.5$ at $\kappa=0$). 
The location of the tip of this cone and the 
detailed shape of its surface are subject to lattice artifacts, as may be the nature 
of the transition for different twist angles.

\vspace{-0.5cm}

\vspace{-0.1cm}
\section{Simulation results}
\label{sec:simulations}
\vspace{-0.1cm}
The simulations were mainly performed on lattices of spatial size $N_{\sigma}=16$ 
and $N_\tau=8$. The algorithm in use is described in Ref.~\cite{Urbach:2005ji}.
An overview of the regions covered 
by our simulation is presented in the left panel of Fig.~\ref{fig:overview}. 
It contains all transition and crossover points that we found in our simulations 
and is supplemented by data describing the $T=0$ critical line 
$\kappa_c(\beta \ge 3.75)$ provided by the ETM collaboration.

After the unsuccessful search reported in~\cite{Ilgenfritz:2007qr},
we have checked the existence of the Aoki phase at the three values 
$\beta\:\in\:\{1.8,\,3.0,\,3.4\}$. The necessary and sufficient condition for 
spontaneous symmetry breaking reads
\begin{equation}
\lim\limits_{h \to 0} \lim\limits_{N_{\sigma}\to\infty} 
\langle \bar{\psi} i \gamma_5 \tau^3 \psi \rangle_{N_{\sigma},\,h} \ne 0 \, .
\end{equation}
\begin{figure}[ht]
\begin{center}
\includegraphics[width=0.48\linewidth]{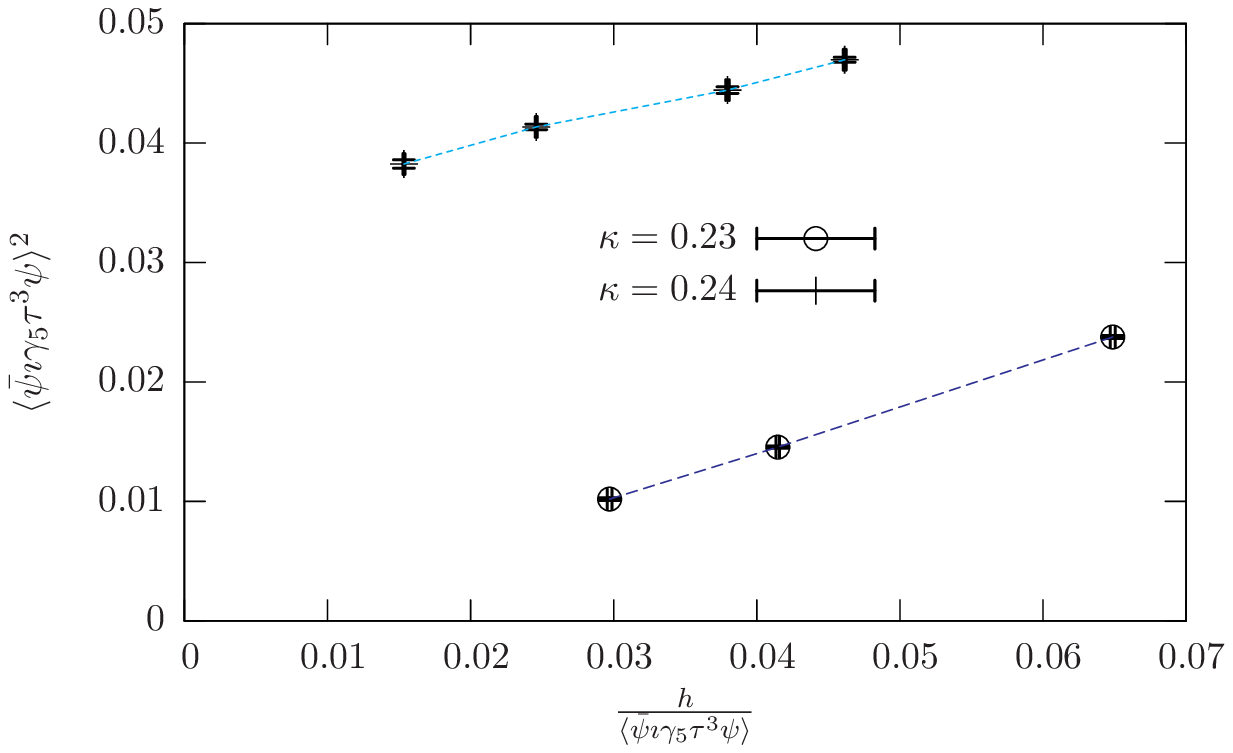}
\includegraphics[width=0.48\linewidth]{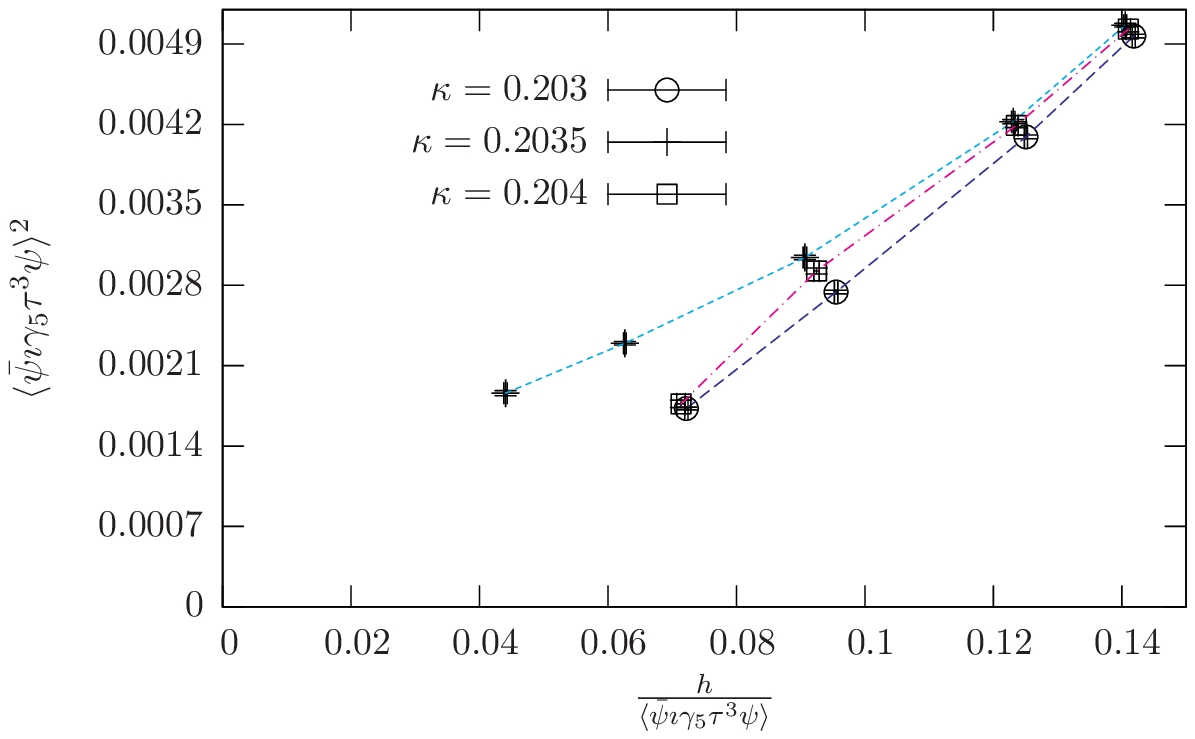}
\end{center}
\vspace{-0.5cm}
\caption{Fisher plots for the order parameter 
$\langle \bar{\psi} i \gamma_5 \tau^3 \psi \rangle$
describing its limit for $h=2\kappa\mu \to 0$; 
{\bf left:} $\beta=1.8$, {\bf right:} $\beta=3.0$ .}
\label{fig:Aokiphase}
\end{figure}
\vspace{-0.1cm}
Figure \ref{fig:Aokiphase} shows the Fisher plots~\cite{Ilgenfritz:2003gw}
of the order parameter $\langle \bar{\psi} i \gamma_5 \tau^3 \psi \rangle$ 
as a function 
of $h/\langle \bar{\psi} i \gamma_5 \tau^3 \psi \rangle$, with $h=2\kappa\mu$, 
for $\beta = 1.8$ and $3.0$. The condensate has a finite limit for $h \to 0$ 
(a positive intercept at $h=0$) for suitable $\kappa$ values.  
For the smallest value $\beta=1.8$, the left panel of
Fig.~\ref{fig:Aokiphase} gives convincing evidence for the onset of 
spontaneous symmetry breaking somewhere 
in the range $0.24 > \kappa > 0.23$. The right panel 
strongly indicates that at $\beta=3.0$ the symmetry-broken phase 
-- if it exists at all -- is realized only in the interval $0.203 < \kappa < 0.204$.
A volume extrapolation would be welcome but is unavailable at this $\beta-$value.
This marks the lower boundary $\kappa_c^{\rm lower}(\beta)$ of the Aoki phase.
We have not systematically searched for its upper boundary 
$\kappa_c^{\rm upper}(\beta)$.

\begin{figure}[h]
\begin{center}
\includegraphics[width=0.49\linewidth]{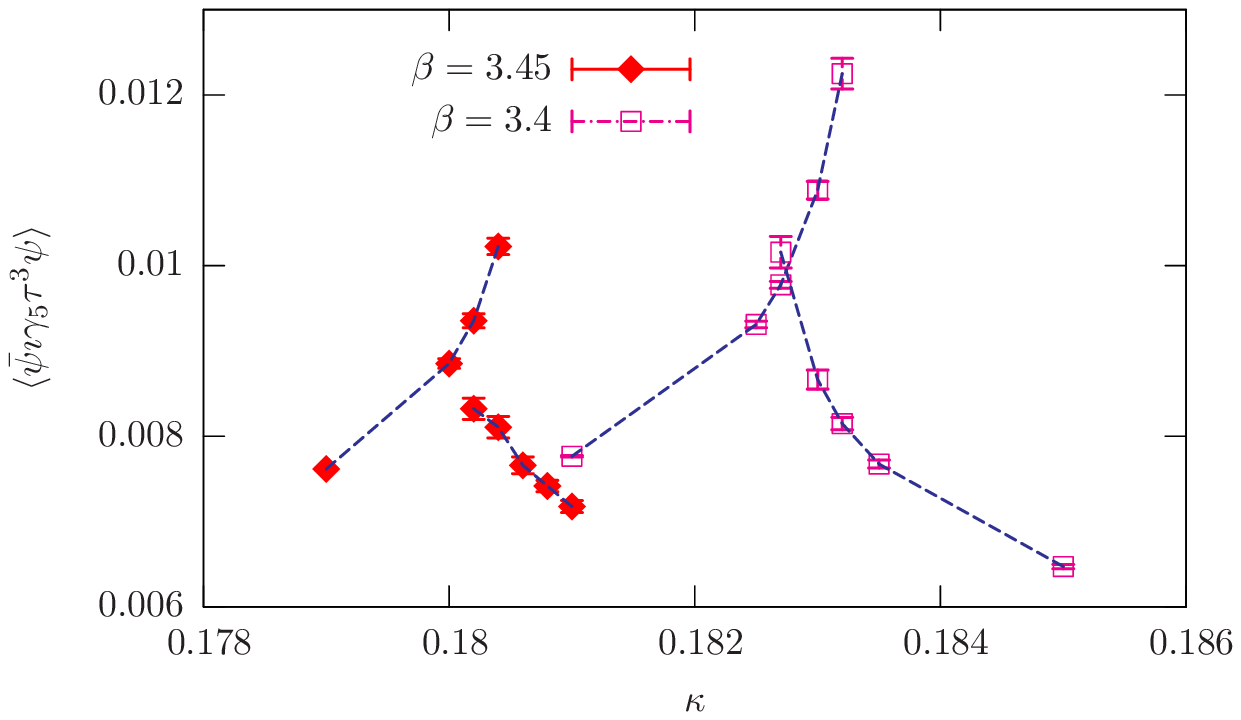}
\includegraphics[width=0.48\linewidth]{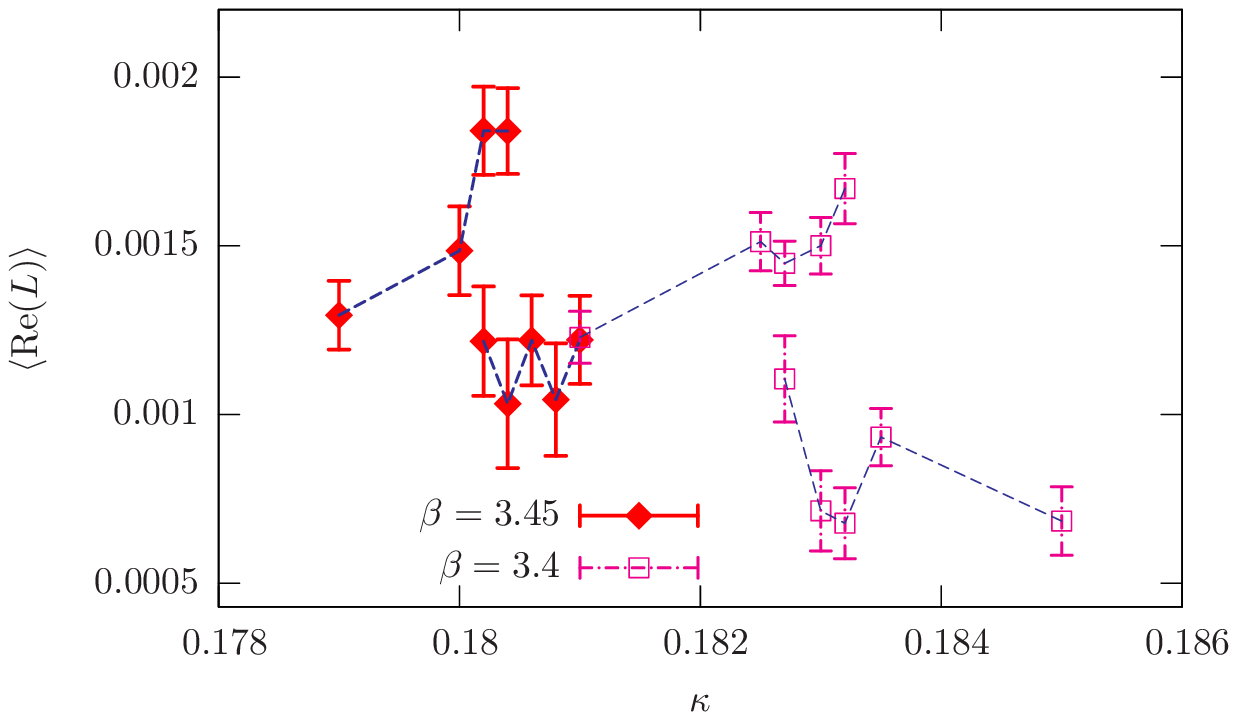}
\end{center}
\vspace{-0.5cm}
\caption{Signals of metastability at $\beta=3.4$ and 3.45.
The two-state signal  
{\bf left:} in the pseudoscalar condensate, 
{\bf right:} in the real part of the Polyakov loop.
\label{fig:metastability}}
\end{figure}
\vspace{-0.2cm}
At $\beta=3.4$ we reinforce our preliminary results published in 
Ref.~\cite{Ilgenfritz:2007qr} that, instead of an Aoki phase, there are strong 
metastabilities when simulating at finite twist $\mu\approx 0.0068$. This confirms the 
first-order Sharpe-Singleton scenario, and represents a remnant of the phase structure 
at $T=0$ (on symmetric lattices). This is shown for $\beta=3.4$ and $3.45$ in 
Fig.~\ref{fig:metastability} for the order parameter (left) and the Polyakov loop
(right). The behaviour of the latter at this 
transition cannot be interpreted as thermal. 
Also the average plaquette (as checked at $\beta=3.4$ and $\kappa=0.1827$)
is running in separate, metastable histories starting from hotter/colder 
neighbouring states.

\begin{figure}[h]
\begin{center}
\includegraphics[width=0.46\linewidth]{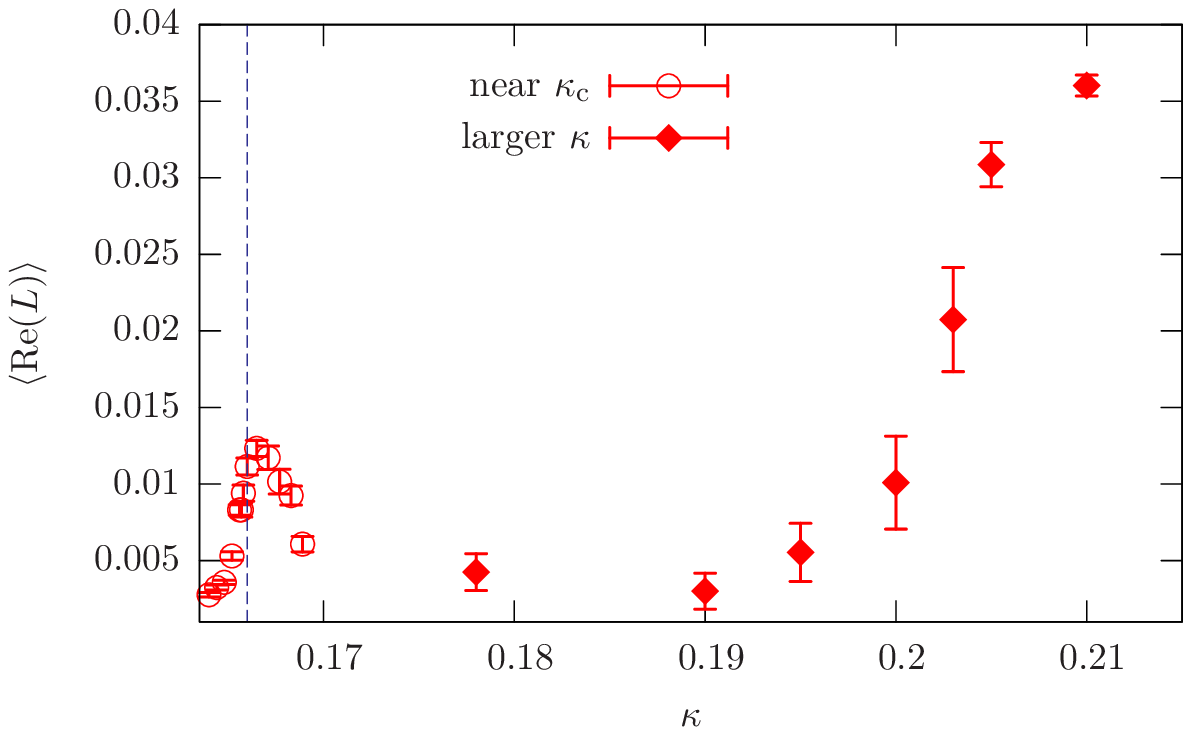}
\includegraphics[width=0.46\linewidth]{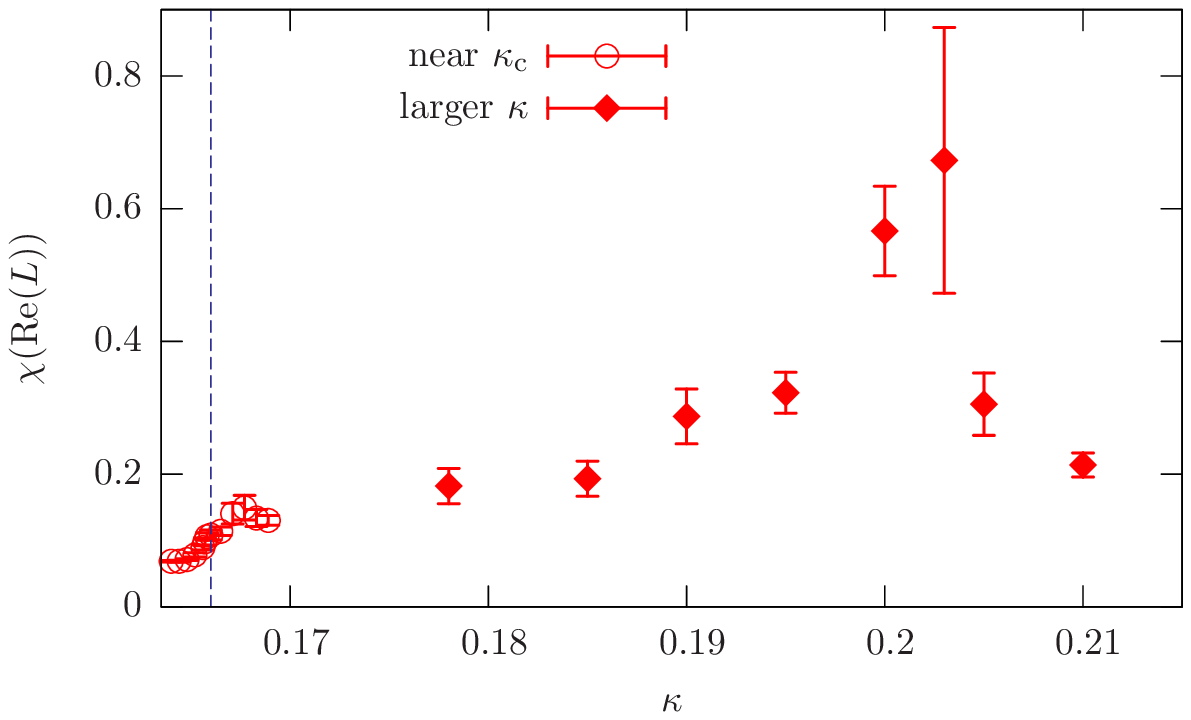}
\end{center}
\vspace{-0.5cm}
\caption{{\bf Left:} real part of the Polyakov loop, {\bf right:} its susceptibility 
as functions of $\kappa$ for $\beta=3.75$ and $\mu=0.005$. The dashed line marks 
$\kappa_c(\beta=3.75,T=0)$.
\label{fig:profile}}
\end{figure}

In the range $3.4 \le \beta \le 3.65$ between strong and weak coupling, 
simulating at $\mu=0.0068$, we see a finite temperature transition, located at 
$\kappa_T \gg \kappa_c(\beta)$, {\it i.e.} far above the chiral line 
for $T=0$. The transition is inherited from the first doubler structure 
existing at higher $\kappa$. The corresponding $\kappa_T(\beta,\mu)$ evolves 
rapidly with $\beta$, moving closer towards the critical line $\kappa_c(\beta)$ 
with increasing $\beta$ (cf. Fig.~\ref{fig:overview}). 
The transition is possibly of first order as indicated by histograms of the 
Polyakov loop at $\beta=3.6$ in the interval $0.18 < \kappa < 0.24$: 
they show a relatively flat but clear double peak structure at $\kappa=0.22$. 
On $32^3 \times 8$ lattices simulated at the same $\beta-\kappa-\mu-$point 
one observes long-living metastability in the Polyakov loop.
Also at higher $\beta=3.75$ this thermal transition continues 
to exist, still separated from the unfolding thermal transition surface around 
$\kappa_c(\beta)$. Finally, it appears to join the latter in the neighbourhood 
of $\beta=4.0$. Before this joining happens, Fig.~\ref{fig:profile} shows the 
characteristic behaviour of the Polyakov loop and its susceptibility measured 
at $\beta=3.75$ with $\mu=0.005$ over an interval $0.16 <\kappa\le 0.21$ that is
covering both transition regions. The two peaks are clearly visible, but the lower 
susceptibility peak is not yet well resolved on the coarse $\kappa$ scale of
Fig.~\ref{fig:profile}.
%% \vspace{-0.1cm}

\begin{figure}[ht]
\begin{minipage}{.45\textwidth}
\hspace{-2mm}\includegraphics[width=\textwidth]{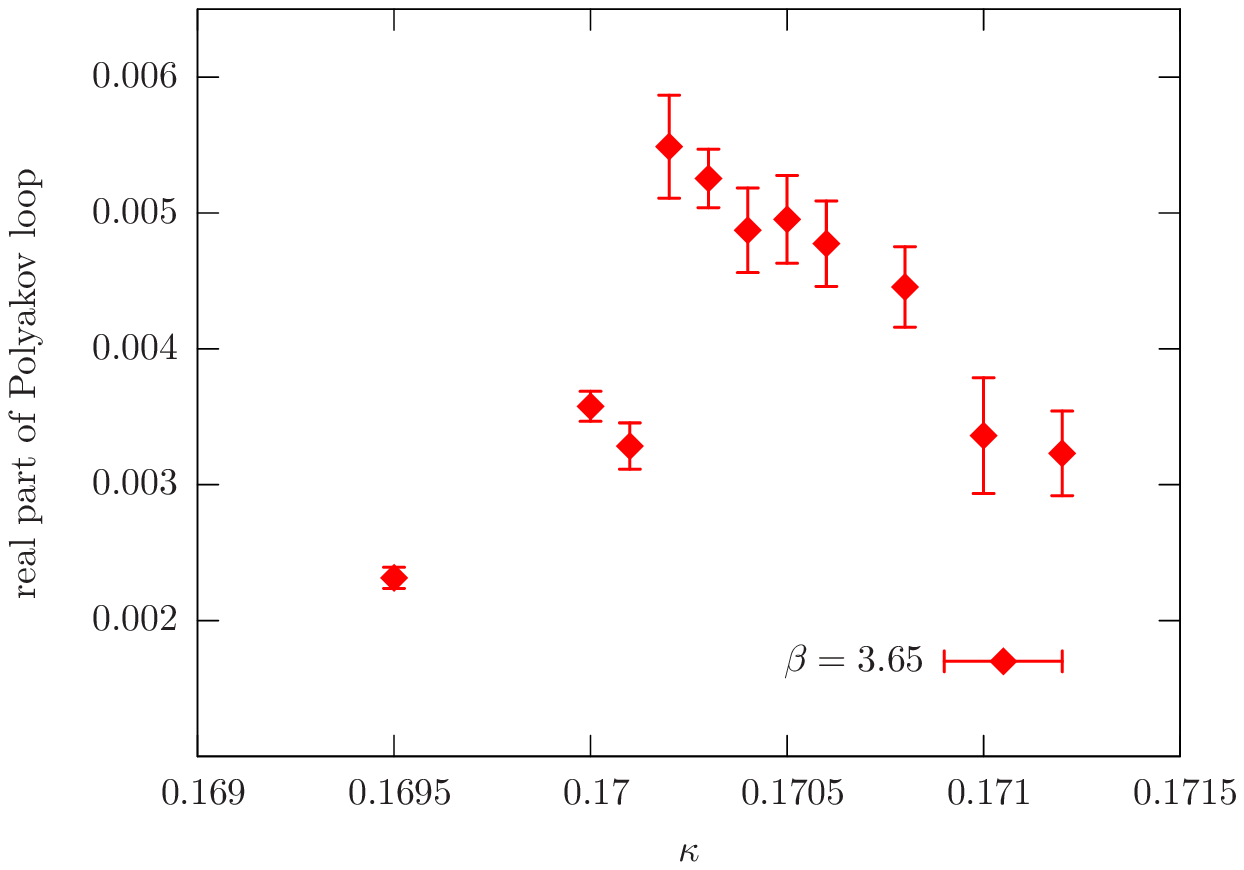}
\end{minipage}\hfill
\begin{minipage}{.45\textwidth}
\includegraphics[width=0.65\textwidth,angle=-90]{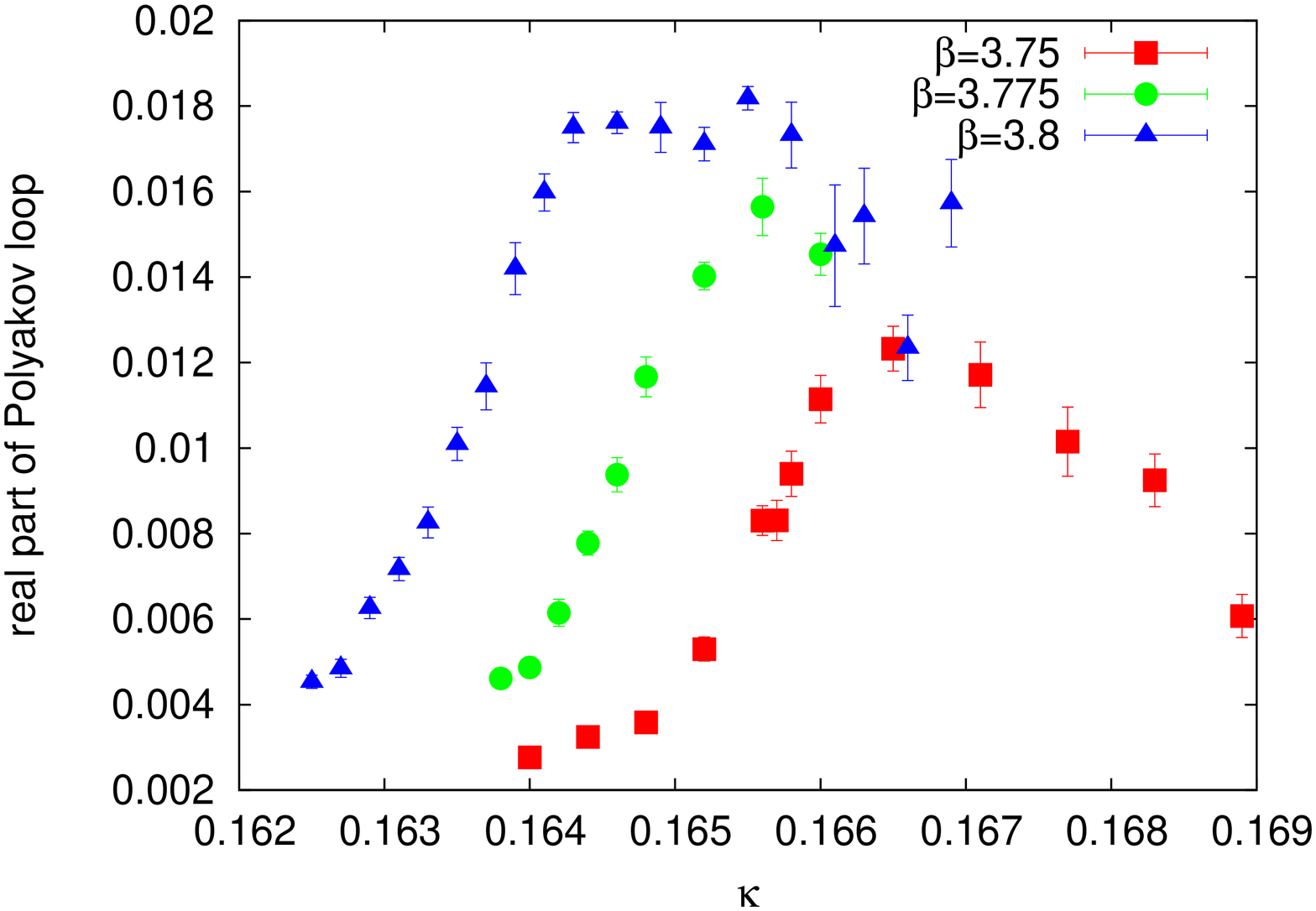}
\end{minipage}
\caption{The real part of the Polyakov loop as function of $\kappa$; {\bf left:} 
for $\beta=3.65$ with $\mu=0.0068$, {\bf right:} for $\beta=3.75$, $3.775$ and 
$3.8$ (all with $\mu=0.005$).
\label{fig:further_change_with_beta}}
\end{figure}

It is our goal to work at maximal twist, thus ultimately we need  
to resolve the lower peak in Fig.~\ref{fig:profile}. The position and nature of 
the first ({\it i.e.} lower in $\kappa$) finite temperature transition/crossover, 
when studied at $\mu \ne 0$, changes considerably between
the metastability region described by Fig.~\ref{fig:metastability} and the 
scaling region. Midway between the two, at $\beta=3.65$, one sees the Polyakov 
loop entering and then leaving again a narrow deconfining 
region in $\kappa$ (cf. the left panel of Fig.~\ref{fig:further_change_with_beta}).
Approaching the physically relevant scaling region, $\beta  \gtrsim 3.75$, the 
peak structure grows even clearer. 
The narrow peak (supposed to contain the deconfined phase) moves gradually further 
to lower $\kappa$ with rising $\beta=3.75$, $3.775$ and $\beta=3.8$
(cf. the right panel of Fig.~\ref{fig:further_change_with_beta}).
\vspace{-0.1cm}

\begin{figure}[h]
\begin{center}
\includegraphics[width=0.32\textwidth,angle=270]{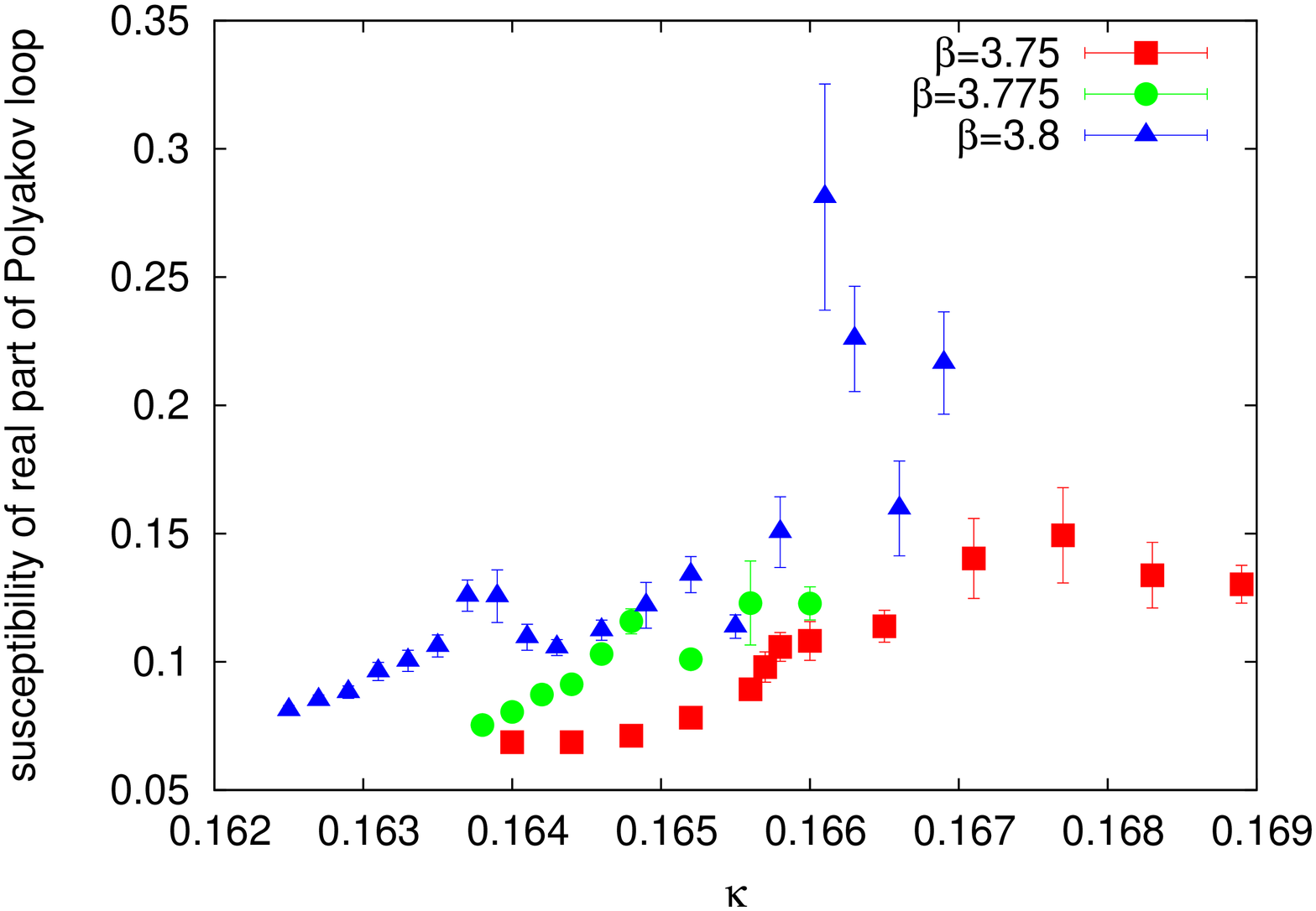}
\hspace{0.5cm}
\includegraphics[width=0.32\textwidth,angle=270]{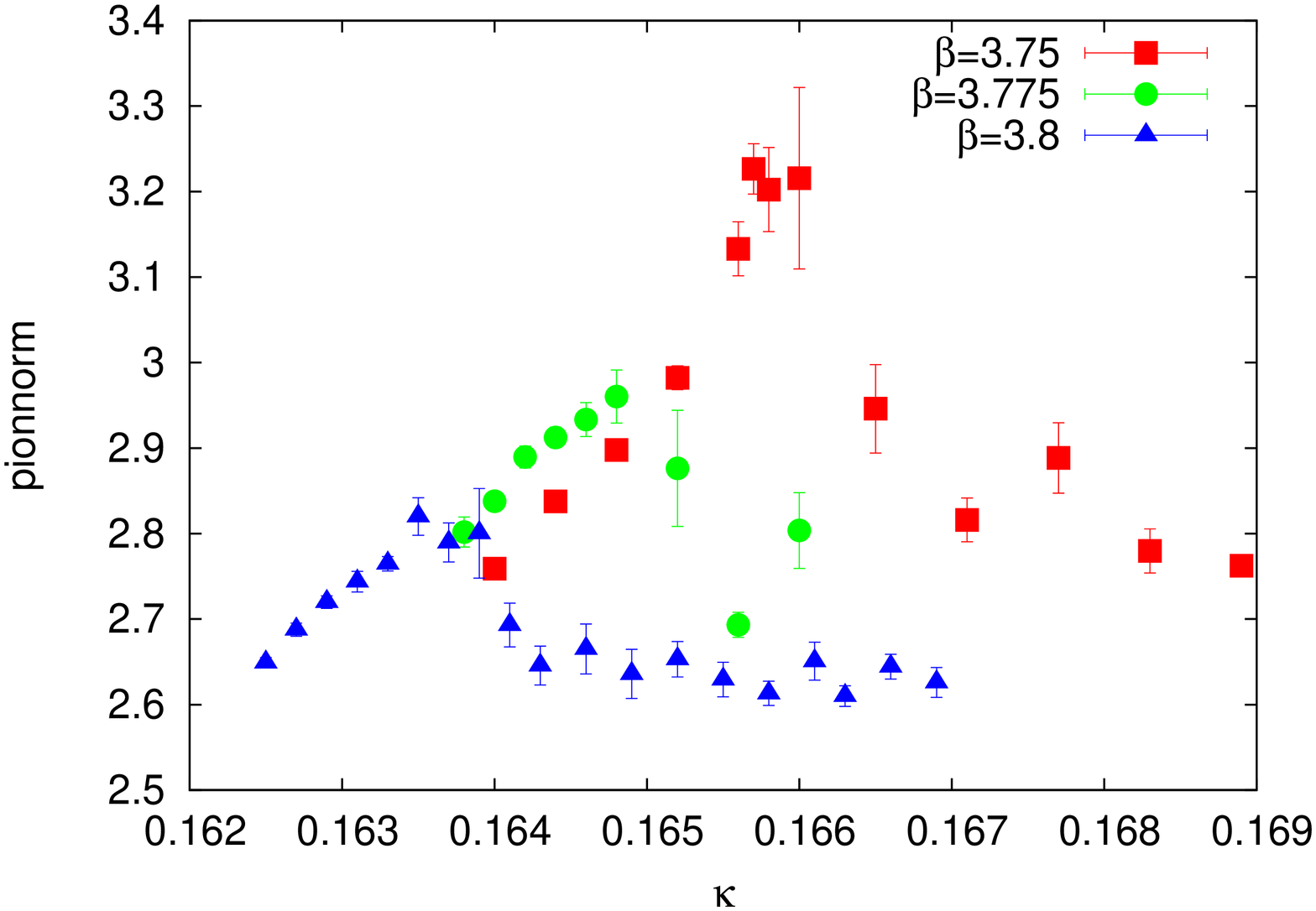}
\end{center}
\vspace{-0.5cm}
\caption{{\bf Left:} the susceptibility of the Polyakov loop, {\bf right:} the 
pion norm as functions of $\kappa$ for $\beta=3.75$, $3.775$ and $3.8$ 
(all at $\mu=0.005$).
\label{fig:chiral_restoration}}
\end{figure}

With a large computational effort the rise and decline of the Polyakov loop begins 
to be reflected also in the corresponding susceptibility. 
The Polyakov loop susceptibility gradually develops a double peak which is
already clearly visible at $\beta=3.8$ 
(cf. the left panel of Fig.~\ref{fig:chiral_restoration}). 
Remarkably, the lower$-\kappa$ peak is accompanied by a peak of the pion norm 
(presented in the right panel of Fig.~\ref{fig:chiral_restoration})
emphasizing the entrance into the deconfining and chiral-symmetry restored 
phase. So far, within the presently available statistics, 
we could not detect a significant similar chiral signal marking the exit. 
\vspace{-0.3cm}

\section{The cone signal and comparison with lattice $\chi{\rm PT}$}
\label{sec:chipt}
\vspace{-0.1cm}
We interpret the structured signal in the Polyakov loop near $\kappa_c$ 
outlined in the preceding section as a numerical indication for passing 
through a confinement$\rightarrow$deconfinement crossover approaching 
$\kappa_c$ from small positive quark mass, followed by a 
deconfinement$\rightarrow$confinement cross\-over at small negative quark mass. 
This picture is qualitatively consistent with the conical structure predicted 
by Creutz (see also Fig.~\ref{fig:overview}).
The thermal transition takes place at a given value of the quark mass which is determined by the parameters $\kappa$ and $\mu$. At tree level the according relation reads:
\begin{equation}
m_q^2 = \mu^2 + \frac{1}{4}\left(\frac{1}{\kappa}-\frac{1}{\kappa_c}\right)^2 \, .
\label{eq:tree}
\end{equation}
\vspace{-0.1cm}
Using lattice chiral perturbation theory (L$\chi{\rm PT}$), one obtains the 
following modification of (\ref{eq:tree}) in NLO (cf.~\cite{Sharpe:2006pu}):
\begin{equation}
m_q^2 = \left(\frac{1}{Z_P^2}\mu^2 + \frac{1}{Z_S^2}\frac{1}{4}\left(\frac{1}{\kappa}-\frac{1}{\kappa_c}\right)^2\right)\left(1+K\cos\omega\right)^2 \, .
\label{eq:chptnlo} 
\end{equation}

From the zero temperature simulations of the ETM collaboration we know that 
$\kappa_c(\beta=3.75)=0.1660(1)$, $Z_S\approx0.6$ and $Z_P\approx0.3$. 
$K$ is an unknown $O(a)$ coefficient that is introduced by L$\chi{\rm PT}$. 
The twist angle $\omega$ is defined by 
$\tan\omega=\mu/\left(0.5\left(1/\kappa-1/\kappa_c\right)\right)$.

\begin{figure}[h]
\centering
\includegraphics[width=0.5\textwidth,height=0.3\textwidth]{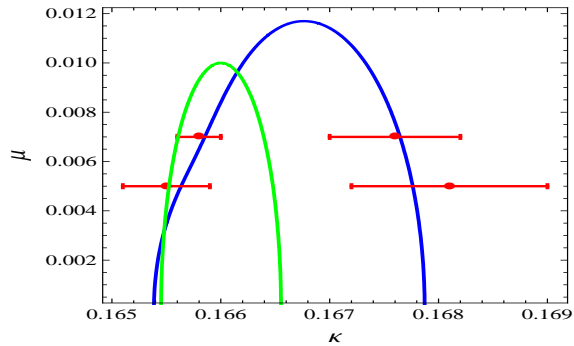}
\caption{Thermal transition points found at $\beta=3.75$ with $\kappa_c=0.1660$ 
compared to tree level ($m_q=0.01$, green) and L$\chi{\rm PT}$ 
($m_q=0.028$, $K=0.5$, $Z_S=0.6$, $Z_P=0.3$, blue) curves.}
\label{fig:cone_plots}
\end{figure}

In Fig.~\ref{fig:cone_plots} the thermal transition points that we 
have found at $\beta=3.75$ for $\mu=0.005$ and $0.007$ are compared 
with the two relations above. 
In principle, $K$ and $m_q$ can be determined from a fit. However,
given the rather large uncertainties of $Z_P$ and $Z_S$ and also the small 
number of available data points, we are only able to check whether the formulae 
are capable of describing the data. For the tree level formula this is not the case 
(in the plot: $m_q=0.01$, $\kappa_c=0.1660$), due to the 
$\kappa\leftrightarrow\left(2/\kappa_c-1/\kappa\right)$-symmetry of 
equation~(\ref{eq:tree}). 
However, choosing $m_q=0.028$, $K=0.5$, $Z_S=0.6$, $Z_P=0.3$ and $\kappa_c=0.1660$ 
the NLO formula~(\ref{eq:chptnlo}) is consistent with the data points.
\vspace{-0.5cm}

\vspace{-0.1cm}
\section{Summary and outlook}
\label{sec:summary}
\vspace{-0.1cm}
We investigated the global phase structure of 2-flavour Wilson-twisted-mass LQCD
with a tree-level Symanzik-improved gauge action. Our simulation results support
the existence of an Aoki phase (at $\beta \lesssim 3.0$) as well as the 
first-order Sharpe-Singleton-scenario at finite twist
(for $\beta \approx 3.4$). Moreover, we find indications 
of a conical structure of the finite-temperature crossover surface in a bounded 
$\beta-$interval, $3.65 \le \beta \le 3.8$, with $\mu=0.005$ and $0.007$.
Our next steps
will be the localization of additional finite temperature crossover 
points to refine the L$\chi{\rm PT}-$prediction of the transition line and to bound
the value $\mu_T$ for the thermal crossover at maximal
twist within a narrow interval. Depending on the findings of this investigation, 
a subsequent simulation 
at maximal twist to detect $\mu_T$ numerically will follow.

\vspace{0.2cm}
{\bf Acknowledgements.} It is a pleasure to thank Steve Sharpe, 
Andr{\'e} Sternbeck and Carsten Urbach
for discussions. In addition, we thank Andr{\'e} for joining our phone conferences
and Carsten for supplying the HMC code. Most simulations have been done at the 
apeNEXT in Rome and DESY-Zeuthen. This work has been supported in part by the DFG 
Sonderforschungsbereich/Transregio SFB/TR9-03. 
O.~P. and L.~Z. acknowledge support by the DFG project PH 158/3-1.
E.-M.~I. was supported by DFG under contract FOR 465 / Mu932/2 
(Forschergruppe Gitter-Hadronen-Ph\"anomenologie). He is grateful to the 
Karl-Franzens-Universit\"at Graz for the guest position he holds while this 
paper is written up.

\end{document}